# Brain Controllability: not a slam dunk yet.

Samir Suweis[1,2,*], Chengyi Tu[3,4], Rodrigo P. Rocha[2,5,6], Sandro Zampieri[2,7], Marzo Zorzi[2,8,9], Maurizio Corbetta[2,10,11]

*[1] Dipartimento di Fisica e Astronomia, 'G. Galilei' & INFN, Università di Padova, Padova, IT. [2] Padova Neuroscience Center, Università di Padova, Padova, IT.[3] Department of Environmental Science, Policy, and Management, University of California, Berkeley, USA. [4] School of Ecology and Environmental Science, Yunnan University, Yunnan, China [5] Department of Physics, School of Philosophy, Sciences and Letters of Ribeirão Preto, University of São Paulo, Ribeirão Preto, SP, Brazil [6]Departamento de Física, Universidade Federal de Santa Catarina, 88040-900, Florianópolis-SC, Brazil [7] Dipartimento di Ingegneria dell'informazione, Università di Padova, Padova, IT. [8]Dipartimento di Psicologia Generale, Università di Padova, Padova, IT. [9] Fondazione Ospedale San Camillo IRCCS, Venezia, IT [10] Dipartimento di Neuroscienze, Universita' di Padova, Padova, IT. [11] Departments of Neurology, Radiology, Neuroscience, and Bioengineering, Washington University, School of Medicine, St. Louis, USA. *samir.suweis@unipd.it*

In our recent article [1] published in this journal we provide quantitative evidence to show that there are warnings and caveats in the way Gu and collaborators [2] define brain controllability.
The comment by Pasqualetti et al. [3] confirms the need to go beyond the methodology and approach presented in Gu et al.'s original work. In fact, they recognize that *"the source of confusion is due to the fact that assessing controllability via numerical analysis typically leads to ill-conditioned problems, and thus often generates results that are difficult to interpret"*. This is indeed the first warning we discussed in [1]: our work was not meant to prove that brain networks are not controllable from one node, rather we wished to highlight that the one node controllability framework and all consequent results were not properly justified based on the methodology presented in Gu et al. [2]. We used in our work the same method of Gu et al. not because we believe it is the best methodology, but because we extensively investigated it with the aim of replicating, testing and extending their results. And the warning and caveats we have proposed are the results of this investigation.
Indeed, after the controllability analyses of multiple human brain networks datasets, we concluded: "*The $\lambda_{min}(W_K)$ are statistically compatible with zero and thus the associated controllability Gramian cannot be inverted[1]. These results show that it is not possible to infer one node controllability of the brain numerically*".

(*Controllability of brain networks from a single region*). We appreciate that Pasqualetti et al., stimulated by our discussion and commentary, developed a new approach [4] - different from the one proposed in Gu et al.- where they show that *a specific* brain network (with no systematic analysis of different networks) is controllable from *a specific node* (*not any node* as they claim in Gu et al. [2]). However, average and modal controllability in [2] are calculated for each node of different brain networks, i.e. the brain can be controlled from any single node of the network, and all conclusions depend on this hypothesis.
We therefore acknowledge that the new method [4] is a potential suitable approach to infer structural controllability from a single node. Nevertheless, we highlight that a systematic test of the methods on all the nodes and the different brain network datasets presented in [2] and related conclusions (beyond one node controllability, e.g. controllability profiles of brain areas) are missing in [4].

(*Controllability of human brain networks versus the C. elegans neuronal network*). Beyond the challenging proof that brain networks may be controllable from every single node, an important theoretical open question is the relationship between network complexity and controllability. In [1] we showed, using a method to detect driver nodes in directed networks [5]), that the control of the C.elegans connectome, which is precisely known, requires about 7% of the nodes. In [3] Pasqualetti et al. argue that C. elegans network control requires more nodes than human networks because they are simpler, and that more complex networks are more easily controllable. In fact, the average connectance (e.g. the density of links) of the C. Elegans and of the different human brain networks we have used in [1] is comparable (0.0715 vs. 0.077, respectively). Therefore, with respect to this feature (connectance), the two types of networks have the same "complexity". Indeed, network complexity is a much more "complex" property than connectance, as it is not simply related to the

---

[1] Finding negative eigenvalues for the Gramian is the consequence of the non-invertibility of the matrix, i.e. when solving the Lyapunov equation to find W, if the linear system is at the edge of instability, then W may have negative eigenvalues. This is in fact also what Gu et al. find, in fact they say in [2]: "*These values (smallest eigenvalues of the Gramian) … remained small (mean $2.5*10^{-23}$, standard deviation $4.8*10^{-23}$)*". This result highlights that also Gu et al. [2] have found negative eigenvalues of W within one standard deviation (i.e. $2.5*10^{-23} -4.8*10^{-23}$ <0)..

number of links in the network. For example, one may consider networks with the same connectance but different architectures. For instance, Erdos-Reny graphs have a "simple" homogenous random structure, while Barabasi-Albert scale free graphs have an heterogenous architecture. Which of the two types of networks is more complex? One can argue that the latter (scale free) are more complex than simple random graph. Yet, it can be proved analytically [5] that directed random Erdos-Reny graphs can be controlled with fewer driver nodes than directed scale free networks. Thus, in this case the more complex network requires a higher number of nodes. In particular in our work we wanted to highlight that the directionality of the network (neglected in [2-4]) plays a crucial role in its controllability profiles

(*Difference in the network models*). In [1] we used a linearization of a Wilson-Cowan model around the quiscient state. In this specific case (linearization around **x***=0) we confirm that our model and the linear model presented in [2] are practically equivalent. As an important sidebar, even though the two models can be practically equivalent, it is quite relevant if the derivation of the model is correct or not. Our model is rigorously derived from linearization procedure of non-linear model describing whole brain activity. How was the linear model of Gu and collaborators derived[2]? There is another important consequence: the linear model *only works* around the linearization point (**x***=0), while in Gu et al. the linear model is considered valid for *any* **x*.** These implications are not discussed in [2], nor in [3] and [4].

*(Controllability as a distinct feature of brain networks).* Pasqualetti et al. [3] claim that in our work we state that "connectivity properties of structural networks estimated from diffusion imaging (DSI/DTI) do not play an important role in brain controllability". We disagree, in fact what we have shown is that using the (unreliable) methodology proposed in [2], random networks and empirical networks show similar controllability profiles. The importance of using null random models was the second central warning in our work. However, Pasqualetti et al. [3] maintain that the controllability profiles of random null models and empirical brain networks are fundamentally different, as shown in their Figure 1. Intriguingly, their Figure 1 is very different from Figures 1-2 in our work [1], in which we found no significant segregation between random and brain networks and a different range in average and modal controllability values, even though– as explained above –the same definition of average and modal controllability was used in the two studies (boundary controllability was not considered, but the results do not change). Thus, why such a big difference in the two results?
Confirming [1] and as explained well in [3], the difference in the linear models used in [1] and in [2] does not explain the difference in the outcome of these simulations.
The crucial point is how null models are generated. Pasqualetti et al. [3] assign the weights of different random networks drawing them "from an empirically-estimated fractional anisotropy distribution". In [1], using standard procedures to build null models in network science [6,7], we use the same weights of the empirical networks. Although several details lack in the explanation of the randomization procedure done in [3], and the data they use are not available – preventing reproducibility – we can safely show how the edge weights have a remarkable impact on the relation between modal and average controllability for different networks. In the top panels we show the results by using a Pareto distribution for the edge weights: in this case we can discriminate among different network structures. However, if we change the parameters of the distribution, we found no difference in controllability between brain and random networks (Figure 1 bottom). This clearly shows that edge weights are a crucial factor discriminating between controllability profiles of different networks, and thus they need to be cautiously and properly discussed. In summary, the procedure for assignment of weights to the random network models proposed by Pasqualetti and collaborators is different from the one proposed by Tu et al., and this explains the different results. The reweighting scheme proposed by Pasqualetti et al. is very unusual in network science, where null models are obtained by keeping the same size and connectivity, and just rewiring the links (null model 1 in [1]), or keeping also the same degree distribution of the data (null model 2 in [1]).

---

[2] In their work, Gu et al. correctly state that "Decades of research demonstrate that neural dynamics are nonlinear… Indeed, ref. [10] proposes a linearized model for the nonlinear neural dynamics described by Wilson Cowan model…linear models of a system accurately approximate nonlinear models in a neighborhood of the operating point". This is exactly the procedure that we employed in [1], while in [2] the statement is not followed by implementation.

(*Theoretical versus practical controllability*). It is crucial to understand, especially for potential users in clinical applications, that only practical controllability matters. In fact, for a structurally controllable network, one node controllability is achievable only if an unrealistically enormous amount of energy is available, when the network dimension is large. This fact has two important consequences that make the conclusions in Gu and collaborators [2] debatable. One regards the relevance of one node controllability of the brain network for potential users in clinical applications. Indeed, the huge amount of energy needed to control the system should have engendered some caution in the interpretation of the results. Single-node controllability is a core concept in the theoretical framework of Gu et al. [2] and given the potential for clinical applications also widely discussed in the paper, and in following ones (e.g., predicting how transcranial magnetic stimulation would affect brain dynamics [8]), it would seem that a more cautionary interpretation of the findings is warranted. Nevertheless, the one-node controllability framework is already making its way into the empirical neuroscience literature. For example, a recent study concluded that controllability (measured using the methods of [2]) modulates the effect of neurostimulation on cognitive performance [9]. It remains unclear how a theoretical construct (one-node controllability) that has little practical significance (because control requires a huge amount of energy) would provide an adequate account of experimental results. The second reason of concern on the conclusions in Gu and collaborators [2] is related to the fact the linear model they propose is, at best, the linearization of a more realistic nonlinear model (see above). Hence, in case of large signals (such as the signal needed for high energy one node control), the two models would behave in completely different manners. In other words, in presence of large signals, the linear and the nonlinear models give, in general, different answers to the same question, such as to be or not one node controllable. Eventually, it is worth highlighting that controllability measures (assuming they are reliable) are meaningful and useful in practice only if they can be disentangled from simpler measures of network structure (i.e., we need to be sure that the empirical effect can only be attributed to controllability; ideally one should assess its effect using other network measures as covariates). In this regard, it is useful to remind the reader that modal and average controllability show near-perfect correlations with node degree (see [1], [2], [9]).

In conclusion, this exchange further highlights the need of carefully assessing ‘warnings and caveats’ of the controllability framework in network neuroscience. We strongly disagree with Pasqualetti et al. conclusions that the argument is settled with regard to single node (Kalman) controllability, especially in regards of using average and modal controllability to asses brain networks controllability profiles [2]. We also show that the relationship between topology and controllability depends on how edges weights are assigned, and null models properly built give the same controllability profiles of real brain networks [1], highlighting how the methodology proposed in [2] on which many other subsequent papers are based is unreliable. Our work is not intended to diminish the potential importance of theoretical tools based on control theory that hopefully will be useful to clinical neuroscientists, and indeed Gu et al have greatly contributed in putting this framework at center stage. However, difficult work lays ahead in bridging theoretical brain controllability and possible applications in translational neuroscience.

## References

[1] Tu, C., et al. "Warnings and caveats in brain controllability." NeuroImage 186 (2018): 83-91.

[2] Gu, Shi, et al. "Controllability of structural brain networks." *Nature communications* 6 (2015): 8414.

[3] Pasqualetti, F. et al., “RE: Warnings and Caveats in Brain Controllability”. NeuroImage 197 (2019) 586–588

[4] Menara, T., Bassett, D.S., Pasqualetti, F., 2017. Structural controllability of symmetric networks. IEEE Trans. Autom. Control (in press).

[5] Liu, Y.-Y., et al. "Controllability of complex networks." *Nature* 473.7346 (2011): 167.

[6] Newman, Mark EJ. "The structure and function of complex networks." *SIAM review* 45.2 (2003): 167-256.

[7] Maslov, S., & Sneppen, K. “Specificity and stability in topology of protein networks”. *Science* (2002), *296*(5569), 910

[8] Muldoon, S.F., et al. "Stimulation-based control of dynamic brain networks." *PLoS computational biology* 12.9 (2016): e1005076.

[9] Medaglia, J. D., et al. "Network Controllability in the Inferior Frontal Gyrus Relates to Controlled Language Variability and Susceptibility to TMS." *Journal of Neuroscience* (2018): 0092-17.

[10] Galán, R. F. "On how network architecture determines the dominant patterns of spontaneous neural activity." *PloS one* 3.5 (2008): e2148.

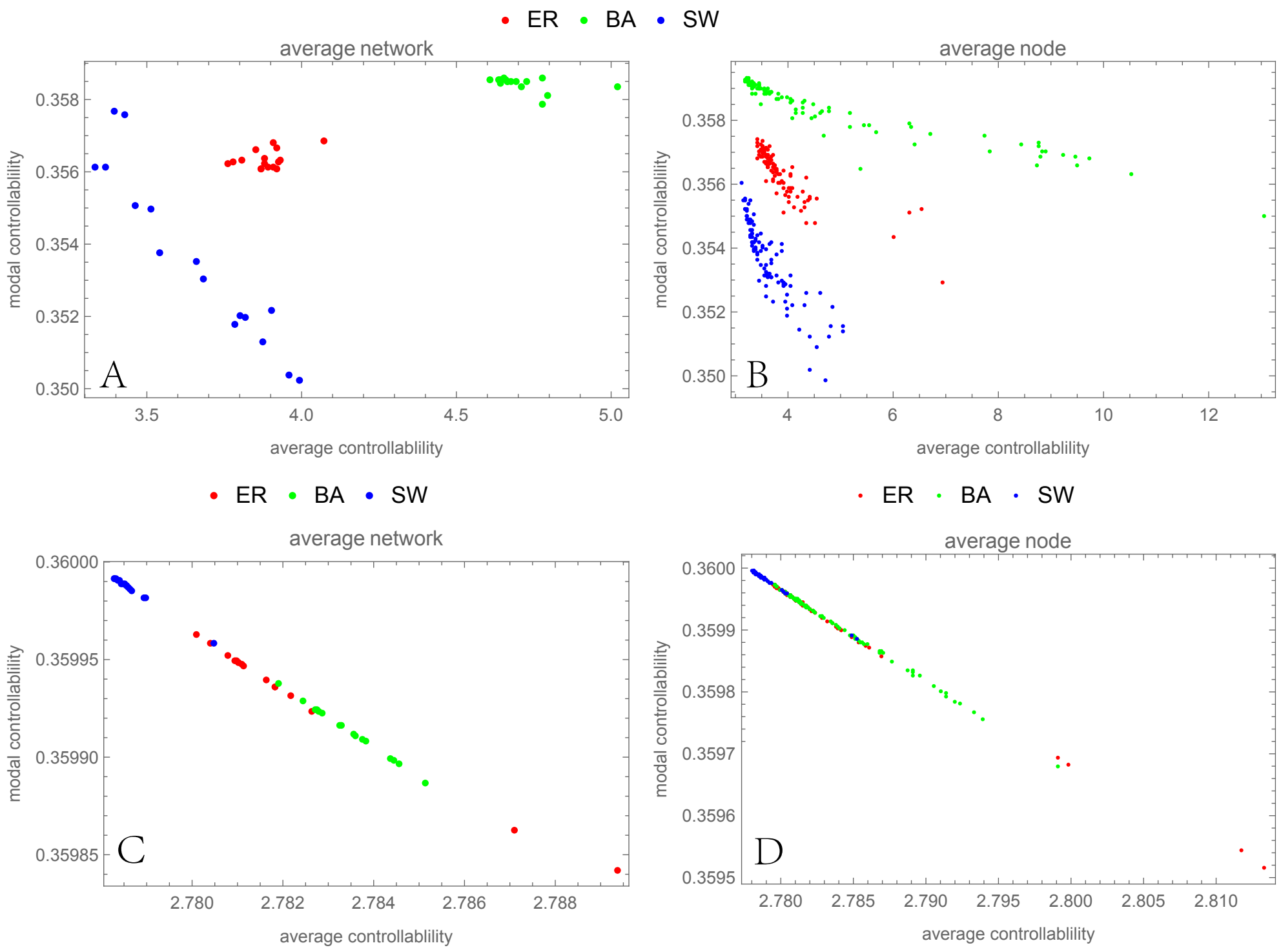

**Figure 1**: Controllability profiles for 100 realizations of random networks (Barabasi-Albert (BA), Small-World network (SW) and Erdős–Rènyi network (ER)) of size 100, connectivity 0.1 and edge weights drawn from Pareto distribution with: Panels A,B) minimum value parameter 2 and shape parameter 3 without normalizing edge weights; Panels C,D) Pareto distribution with minimum value parameter 0.005 and shape parameter 2 after normalization of edge weights). Left panels represent average values computed over all possible control nodes, while in the right panels we report the average values computed over all network instances.